\documentclass[floatfix,twocolumn,aps,prl,reprint,superscriptaddress]{revtex4-1}
\usepackage{amsmath}
\usepackage{amssymb}
\usepackage{float}
\usepackage{graphicx}
\usepackage{color,soul}
\usepackage{hyperref}
\usepackage[colorinlistoftodos]{todonotes}
\usepackage{lipsum}
\usepackage{braket}

\graphicspath{ {figures/} {figures_supplement/} }

\makeatletter

\@ifundefined{textcolor}{}
{%
 \definecolor{BLACK}{gray}{0}
 \definecolor{WHITE}{gray}{1}
 \definecolor{RED}{rgb}{1,0,0}
 \definecolor{GREEN}{rgb}{0,1,0}
 \definecolor{BLUE}{rgb}{0,0,1}
 \definecolor{CYAN}{cmyk}{1,0,0,0}
 \definecolor{MAGENTA}{cmyk}{0,1,0,0}
 \definecolor{YELLOW}{cmyk}{0,0,1,0}
}
\usepackage{dcolumn}
\usepackage{bm}
\usepackage{float}

\usepackage{graphicx}
\makeatother

\makeatother

\begin{document}

\title{
Seamless high-Q microwave cavities for multimode circuit QED
}
\author{Srivatsan Chakram}
\thanks{These authors contributed equally to this work.}
\affiliation{James Franck Institute, University of Chicago, Chicago, Illinois 60637, USA}
\affiliation{Department of Physics, University of Chicago, Chicago, Illinois 60637, USA}
\affiliation{Department of Physics and Astronomy, Rutgers University, Piscataway, New Jersey 08854, USA}

\author{Andrew E. Oriani}
\thanks{These authors contributed equally to this work.}
\affiliation{James Franck Institute, University of Chicago, Chicago, Illinois 60637, USA}
\affiliation{Pritzker School of Molecular Engineering, University of Chicago, Chicago, Illinois 60637, USA}

\author{Ravi K. Naik}
\affiliation{James Franck Institute, University of Chicago, Chicago, Illinois 60637, USA}
\affiliation{Department of Physics, University of Chicago, Chicago, Illinois 60637, USA}
\affiliation{Department of Physics, University of California Berkeley, California 94720, USA}

\author{Akash V. Dixit}
\affiliation{James Franck Institute, University of Chicago, Chicago, Illinois 60637, USA}
\affiliation{Department of Physics, University of Chicago, Chicago, Illinois 60637, USA}

\author{Kevin He}
\affiliation{James Franck Institute, University of Chicago, Chicago, Illinois 60637, USA}
\affiliation{Department of Physics, University of Chicago, Chicago, Illinois 60637, USA}

\author{Ankur Agrawal}
\affiliation{James Franck Institute, University of Chicago, Chicago, Illinois 60637, USA}
\affiliation{Department of Physics, University of Chicago, Chicago, Illinois 60637, USA}

\author{Hyeokshin Kwon}
\affiliation{Samsung Advanced Institute of Technology (SAIT), Samsung Electronics Co., Ltd, Suwon, Republic of Korea}

\author{David I. Schuster}
\email{David.Schuster@uchicago.edu}
\affiliation{James Franck Institute, University of Chicago, Chicago, Illinois 60637, USA}
\affiliation{Department of Physics, University of Chicago, Chicago, Illinois 60637, USA}
\affiliation{Pritzker School of Molecular Engineering, University of Chicago, Chicago, Illinois 60637, USA}

\begin{abstract}
Multimode cavity quantum electrodynamics ---where a two-level system interacts simultaneously with many cavity modes---provides a versatile framework for quantum information processing and quantum optics. Due to the combination of long coherence times and large interaction strengths, one of the leading experimental platforms for cavity QED involves coupling a superconducting circuit to a 3D microwave cavity. In this work, we realize a 3D multimode circuit QED system with single photon lifetimes of $2$ ms and cooperativities of $0.5-1.5\times10^9$ across 9 modes of a novel seamless cavity. We demonstrate a variety of protocols for universal single-mode quantum control applicable across all cavity modes, using only a single drive line. We achieve this by developing a straightforward \textit{flute} method for creating monolithic superconducting microwave cavities that reduces loss while simultaneously allowing control of the mode spectrum and mode-qubit interaction. We highlight the flexibility and ease of implementation of this technique by using it to fabricate a variety of 3D cavity geometries, providing a template for engineering multimode quantum systems with exceptionally low dissipation. This work is an important step towards realizing hardware efficient random access quantum memories and processors, and for exploring quantum many-body physics with photons.
\end{abstract}

\maketitle

\begin{figure}[t]
\includegraphics[width=.5\textwidth]{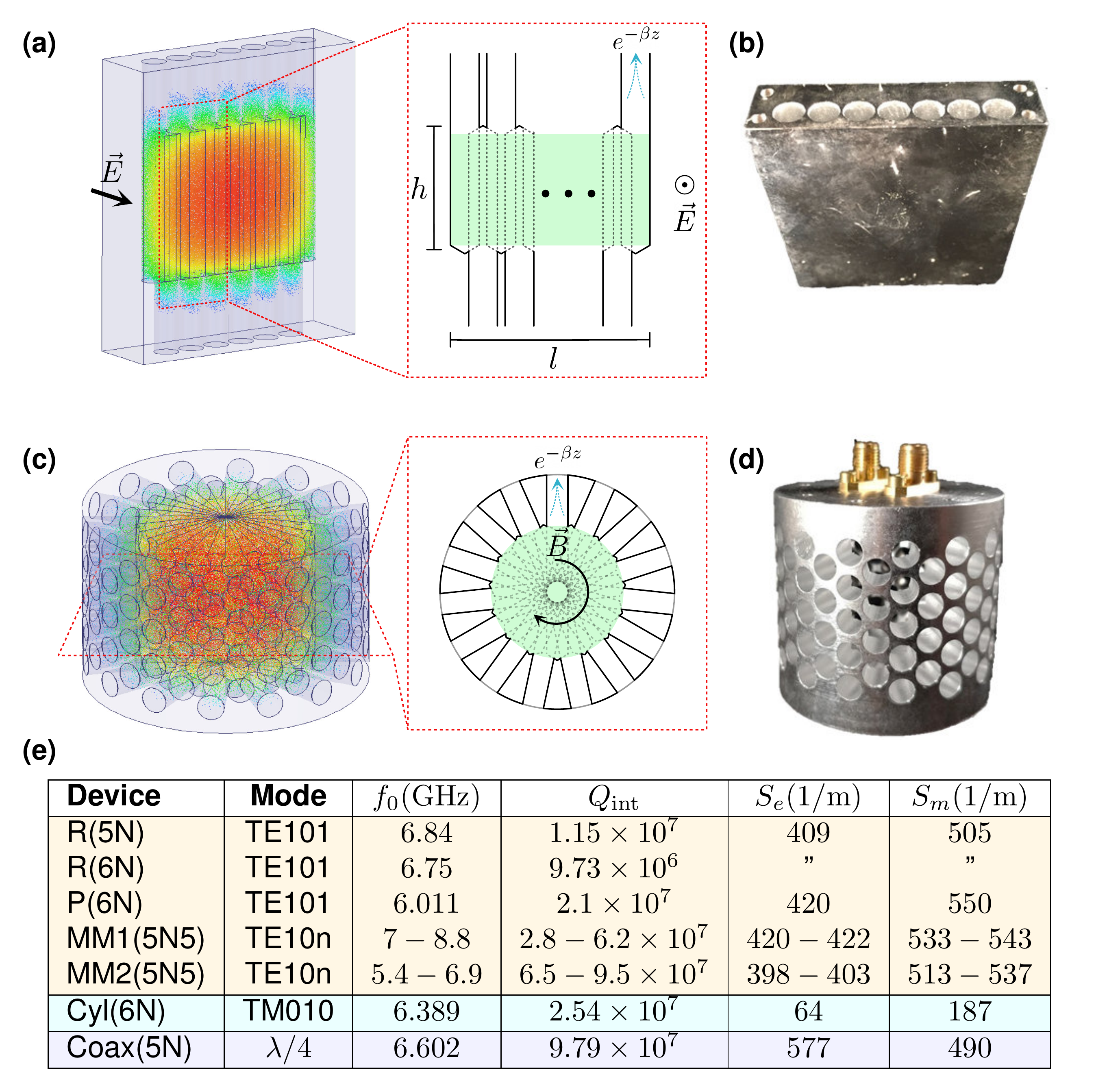}
\caption{\textbf{Outline of seamless \textit{flute} cavity design}. (a) An FE model showing the $\vec{E}$-field magnitude for the $\mathrm{TE_{101}}$ mode of a rectangular waveguide cavity. (inset) A side-view cutaway of the flute design highlighting the overlapping holes, with the effective mode volume highlighted in green. The evanescent decay through the holes is also shown. (b) A picture of the R(5N) cavity. (c) An FE model of a cylindrical style flute cavity showing the $\vec{E}$-field magnitude for the fundamental $\mathrm{TM_{010}}$ mode. (inset) A top-view cutaway showing the effective mode volume created by the hole overlap. (d) A picture of the C(6N) cavity. (e) A table outlining the performance of various cavity geometries, highlighting the internal quality factors ($Q_{\mathrm{int}}$), and the magnetic ($S_m$) and electric ($S_e$) participation ratios from FE simulations. }

\label{Fig1}
\end{figure}

Circuit quantum electrodynamics (cQED)~\cite{wallraff2004strong} has emerged as the preeminent platform for quantum optics and realizing quantum memories~\cite{reagor2016superconducting}. While studies of quantum optics with cQED have largely been restricted to a single or few cavity modes, the extension of cQED to many cavity modes (multimode cQED)
promises explorations of many-body physics with exquisite single photon control. Multimode cavities are an efficient way of realizing many co-located cavity modes that can be simultaneously coupled to and controlled by a single physical qubit, ideal for creating multi-qubit quantum memories while reducing the number of physical lines required.  A challenge currently limiting applications of multimode cQED in quantum information science---and the scaling to larger systems---is the need for longer coherence times. 

3D superconducting cavities possess the longest coherence times in cQED~\cite{Reagor2013}, and while being intrinsically linear, can be strongly coupled to a nonlinear superconducting transmon circuit to realize universal gate operations~\cite{heeres2015cavity, heeres2017implementing}. The resulting high cooperativities have enabled many fundamental experiments in quantum information science and quantum optics, including demonstrations of quantum error correction~\cite{ofek2016extending, hu2019quantum, campagne2020quantum} and fault tolerance~\cite{reinhold2020error}. While quantum control has also been extended to two-cavity modes coupled to the same qubit~\cite{wang2016schrodinger}, where it has been used to mediate gate operations and interactions~\cite{rosenblum2018cnot,gao2019entanglement}, it has so far not been extended to many cavity modes. Building multimode systems that leverage 3D cavities will enable explorations of a new regime of many-body quantum optics. Using a single physical qubit to control a multimode memory also allows us to potentially multiplex $\sim10-1000$ modes, thereby providing a promising solution to the problem of wiring large quantum processors, and allowing superconducting quantum systems to go beyond the noisy intermediate-scale quantum era~\cite{arute2019quantum, jurcevic2020demonstration}.

Multimode cQED systems with strong light-matter interactions have been realized in a variety of 2D quantum circuits, with a Josephson-junction-based superconducting qubit coupled to many nearly harmonic modes. These include transmission line resonators~\cite{sundaresan2015beyond}, superconducting lumped-element~\cite{mirhosseini2018superconducting, kim2020quantum, indrajeet2020coupling} and Josephson-junction-based meta-materials~\cite{martinez2019tunable}, and electromechanical systems~\cite{pechal2018superconducting}, highlighting the breadth of quantum optics and simulation problems that can be addressed with multimode cQED~\cite{leppakangas2018quantum}. A multimode cQED system comprising a chain of strongly coupled coplanar waveguide resonators has also been used to realize a random access quantum processor in which a single transmon mediates gate operations between arbitrary mode pairs~\cite{naik2017random}. For scaling such multiplexed systems to larger Hilbert spaces, the harmonic modes (quantum memories) must have much longer coherence times than the transmon qubit (quantum bus).

In this letter, we demonstrate a flexible 3D multimode cavity platform capable of high cooperativities across many cavity modes. To do this, we develop a new \textit{flute} technique that enables the creation of a variety of cavity geometries while eliminating seam loss---arising from supercurrents crossing mechanical interfaces---present in the construction of many cavity designs. Using this technique, we realize a state-of-the-art multimode cQED system consisting of a monolithic 3D multimode cavity with coherence times exceeding $2$ ms across the mode spectrum. We perform quantum operations on 9 of the cavity modes using a single superconducting transmon circuit placed at one end of the cavity. We extend a variety of universal cavity control schemes---based on the dispersive interaction---to a multimode system, realizing a promising and flexible platform for hardware efficient random access quantum memories.

The flute technique creates a cavity through the overlap of holes drilled from the top and bottom of a monolithic piece of superconductor, resulting in the generation of a cavity volume with no seams. This is illustrated in Fig.~\ref{Fig1} (a) and (c) for a rectangular and cylindrical cavity, where we show finite element (FE) simulations of the fundamental cavity mode and visualize the creation of the cavity volume through the overlap of the holes (insets, cavity volume highlighted in green). The hole diameter is chosen such that the cutoff frequency of the waveguide mode is much higher than that of the cavity modes. This ensures that the cavity mode energy density at the vacuum interface is suppressed by a factor of $\exp(-\beta z)$, where $\beta$ is the waveguide propagation constant for the $\mathrm{TM_{0m}}$ modes of the hole, and $z$ is the hole depth (from the top of the cavity volume, green). 
We choose a hole depth ($z = 15$ mm) and diameter ($d_o = 4.76$ mm) such that the quality factor limit due to evanescent coupling to vacuum exceeds $10^9$ for all the manipulable modes of the cavity (see SI),  similar to the values for evanescent loss for a single mode coaxial $\lambda/4$ cavity~\cite{reagor2016superconducting}. As a result, despite the cavity being full of holes, we realize high quality factor cavities limited only by intrinsic losses.

We used the flute method to construct a number of cavities, with various geometries, using aluminum ranging in purity from $5$N ($99.999\%$) to $6$N ($99.9999\%$). These include the rectangular \textit{pan-pipe} [R(5N), R(6N)] and cylindrical cavities~[Cyl(6N)] depicted in Fig.~\ref{Fig1} (b) and (d) respectively. Their quality factors, inferred from their spectroscopic linewidths, are shown in the table in Fig~\ref{Fig1} (e), along with their mode frequencies and their geometric surface participation factors. We also measured the coherence of a \textit{pan-pipe} cavity [P(6N)]~\cite{dixit2020searching} and multimode cavity [MM2(5N5)] while coupled to a qubit. During fabrication, care was taken to ensure a smooth and uniform finish of the cavity interior and to reduce the formation of metal burrs at the hole-cavity interface. To do this, successive drilling and honing steps were used to bring the holes to their final dimensions, followed by etching to remove surface damage induced from the manufacturing process (see SI). For comparison, we also measured the internal quality factor of a 5N aluminum coaxial $\lambda/4$ cavity that underwent the same etching process, which had an internal quality factor of $~97$ million---comparable with the best quality factors observed in this cavity geometry in aluminum~\cite{kudra2020high}. When the cavity losses are scaled with respect to their geometric magnetic ($S_{m}$) and electric ($S_e$) surface participation ratios~\cite{Reagor2013,pozar2009microwave}, the MM2(5N5) cavity internal quality factors are comparable to those achieved in the coaxial cavity, ranging from $65-95$ million over the first 9 modes. The losses seen in the other cavity geometries differ by nearly a factor of 2 from that expected from the coaxial cavity Q, even once the geometric scaling is taken into account. The source of this additional non-radiative loss is attributed to variations introduced in the manufacturing and surface treatment (see SI).

\begin{figure}[t]
\includegraphics[width=0.5\textwidth]{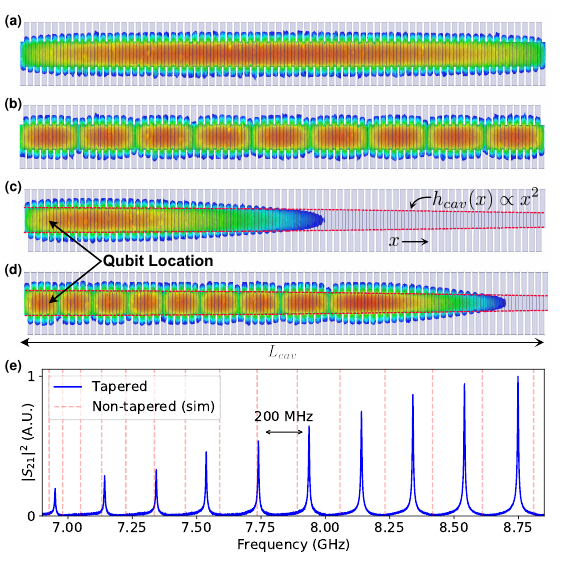}
\caption{\textbf{Dispersion engineering in a multimode flute cavity}
(a),(b) FE simulations showing the magnitude of the electric field ($|\vec{E}|$) for the first, and the ninth mode of a long rectangular cavity. (c),(d) $|\vec{E}|$ profile for a similar cavity with the height tapered according to the expression $h_{\mathrm{cav}}(x) = h_0 - \alpha x^2$. (e) Mode spectrum of a tapered multimode flute cavity [MM1(5N5)] measured at room temperature (blue), and the simulated eigenfrequencies for a non-tapered cavity (red vertical lines) of the same length.}
\label{Fig2}
\end{figure}

\begin{figure*}[t]
\includegraphics[width=1\textwidth]{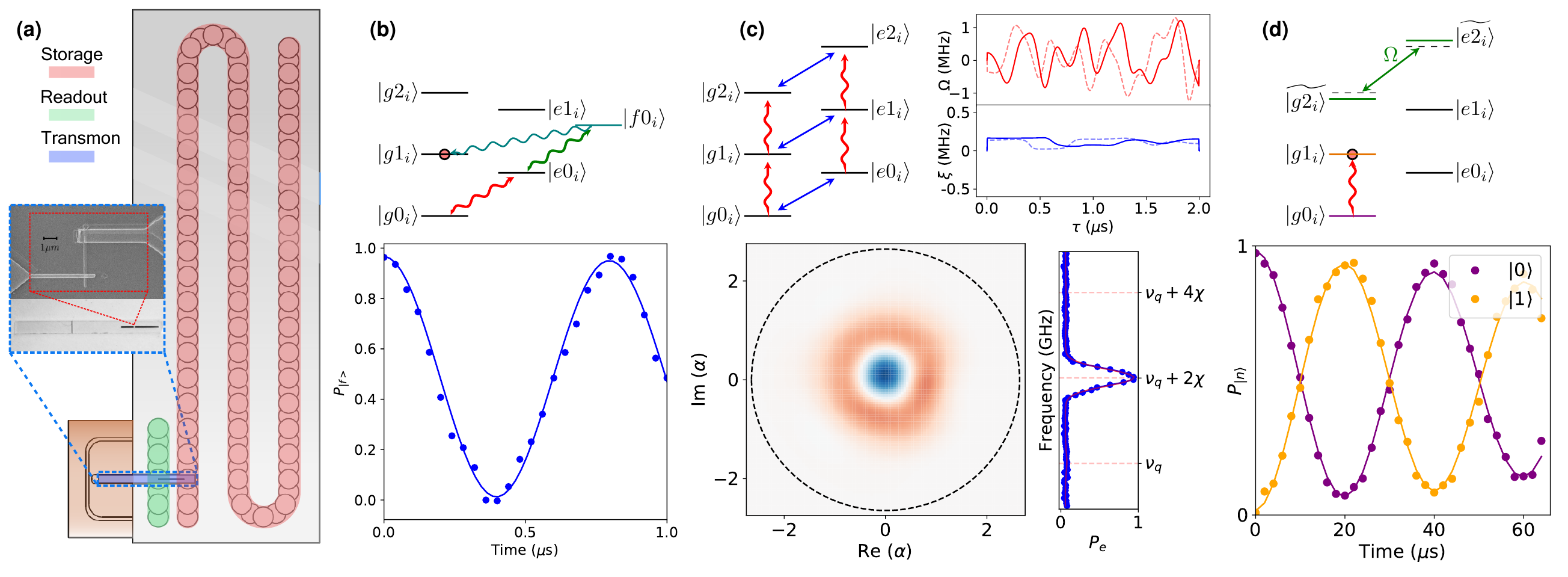}
\caption{\textbf{Quantum control of multimode flute cavity using a transmon}. (a) A schematic of the multimode flute cavity [MM2(5N5)] showing the location of the storage cavity (red), readout cavity (green), and transmon chip (blue). (b, top) Energy level diagram illustrating cavity state preparation using $\ket{f0_i}-\ket{g1_i}$ charge sideband transitions. (bottom) Corresponding sideband Rabi oscillations for mode 3 obtained after initializing the transmon in the $\ket{f}$ state, and driving at the $\ket{f0_3}-\ket{g1_3}$ difference frequency. (c, top) Energy level diagram illustrating control of the cavity using SNAP gates with resonant drives on the cavity (red) and the transmon (blue). (c, top, right) Optimal control pulses on the transmon ($\Omega$) and the cavity ($\xi$), required for preparing cavity mode $2$ in $\ket{1}$. (c, bottom) Measurement of the resulting state using Wigner tomography (left), and photon number resolved qubit spectroscopy (right). (d, top) Energy level diagram illustrating state preparation by photon blockade using a resolved transmon pulse resonant with $\ket{g2_i}-\ket{e2_i}$ (green). The resulting Rabi splitting  makes the cavity mode anharmonic, with a weak resonant cavity drive (red) producing Rabi oscillations, as shown in (d, bottom) for mode 3.}
\label{Fig3}
\end{figure*}

While all 3D cavities are naturally multimodal, the usability of the modes depends on the mode frequencies and electric field participations at the qubit location. We achieve multimode cavities that satisfy both these requirements by using the $\mathrm{TE_{10m}}$ modes of a long rectangular waveguide cavity, as shown in Fig.~\ref{Fig2}. The frequency of the fundamental mode (Fig.~\ref{Fig2} (a)) is tuned by the cavity height ($h$), the second smallest cavity dimension. The cavity spectrum is given by $\nu_{nm} = \frac{c}{2}\sqrt{(n/h)^2+(m/l)^{2}}$, where $m$ and $n$ are the mode numbers along the cavity length and height.
In our case, we operate on modes with a single antinode along the height, such that $n=1$. Higher m-modes have increasing number of antinodes along the length, as illustrated by the $9^{\mathrm{th}}$ mode shown in Fig.~\ref{Fig2}(b).
In this regime, the mode spacing scales inversely with length, with the modes near the cutoff frequency having significant dispersion, as shown in Fig~\ref{Fig2}(e)(dashed red lines). 
We can change this dispersion by modulating the cavity height by varying the top and bottom hole overlap across the cavity length. With a quadratic profile, $h(x) = h_0 - \alpha x^2$, as shown in Fig~\ref{Fig2}(c) and (d) we are able to create a nearly constant mode spacing, as shown by the transmission measurement in Fig.~\ref{Fig2}(e)(blue) for MM1(5N5). This has the additional effect of lensing the field towards one side of the cavity, increasing the coupling to the qubit (location indicated by arrows) for the higher order modes. By modulating the cavity height, we are therefore able to tune both the mode dispersion and mode-qubit coupling, allowing us to tailor the parameters of the mutimode cavity-QED system depending on the application. Increasing the cavity length leads to reduced mode spacing and more modes in a given bandwidth, with the spacing between modes being ultimately limited by off-resonant interactions between the qubit and non-target modes. To reduce these effects, we conservatively chose a mode spacing of $200-250$ MHz. Future devices could be longer and include more modes within the available bandwidth.

We control the cavity modes using a superconducting transmon circuit that serves as a quantum bus that couples to all the modes. The choice of the number of modes is also informed by the coherence times of the transmon and the cavity modes. In a multiplexed random access architecture, the gate error while operating on a target mode should be comparable to the accumulated idle errors of the non-target modes. This suggests that the number of usable modes scales as $n_{m} = T_{1}^{c}/T_{1}^{q}$, where $T_{1}^{c}$ and $T_{1}^{q}$ are the cavity and qubit relaxation times respectively. With our average measured $T^{c}_{1}\sim~2$ ms and $T^{q}_{1}\sim80-100~\mu s$, we are able to control Hilbert spaces of $\sim10$ modes before being limited by deleterious effects due to multiplexing. The superconducting transmon circuit is simultaneously coupled to all the cavity modes by placing it at one end of the multimode flute cavity as shown in Fig.~\ref{Fig3} (a), where the first 9 modes have couplings ranging from $50-170$ MHz. The capacitor pads of the transmon act as antennas that couple to the electric fields of the modes of the storage cavity (red) and a second adjacent smaller flute cavity used for readout (green). This interaction allows the cavity control operations developed in single-mode systems to be applied to any mode of the multimode cavity, all through a single drive line that couples directly to the readout resonator. This corresponds to a ten-fold reduction in the number of control lines compared to a similar sized physical qubit system, and demonstrates the hardware efficiency of this design. By leveraging current niobium accelerator cavity technology, it is possible to realize single photon lifetimes $>2$s~\cite{romanenko2020three} and thereby allowing for multliplexing over $\sim1000$ modes---with qubit lifetimes of $100-200~\mu s$---without needing to add any more control lines.

We demonstrate 3 different ways of controlling the cavity modes, all of which use the Josephson non-linearity of the transmon to exploit different physics of the system. These protocols differ in the required drive strengths, frequencies, and gates times, but result in similar infidelities up to prefactors in the regime where the transmon is the dominant source of decoherence. These are (1) resonant photon exchange mediated by 4-wave mixing processes, (2) cavity displacements used in conjunction with photon number selective phase gates (SNAP)~\cite{heeres2015cavity}, and (3) cavity drives within subspaces engineered by photon blockade~\cite{bretheau2015quantum,chakramhe2020}.
These schemes can also be extended to perform gate operations and generate interactions between modes. The control methods can all be understood by rewriting the junction phase in terms of the dressed states arising from the interaction with the modes, and expanding the transmon Josephson energy to quartic order:
  \begin{equation}
\begin{split}
H_I = \frac{\alpha}{12}\left(\beta_t\hat{c} + \beta_r\hat{a}_r +  \sum_m\beta_m\hat{a}_m + \beta_r\xi_d +  \mathrm{c.c} \right)^4
\label{eqn1}
\end{split}
\end{equation}
Here, $\beta_t, \beta_r, \beta_m$ are the participations of the transmon, readout, and storage modes in the phase of the transmon junction, respectively.  $\xi_d$ is the combined readout and transmon drive displacement precessing at the drive frequency, and all operators are rotating at their natural frequencies (see SI).
This interaction leads to a 4-wave mixing process ($\sim\sqrt{\chi_m\chi_r}\hat{c}^2\hat{a}^{\dagger}_m\xi_d$/2) that takes two photons in the transmon ($\ket{f0_m}$) to one photon in the storage mode ($\ket{g1_m}$) using a single-drive tone at their difference frequency, as illustrated in Fig.~\ref{Fig3}(b). This $\ket{f0}-\ket{g1}$ sideband (\cite{pechal2014microwave, rosenblum2018cnot}) can be used to perform SWAP operations on the modes in $0.5-1~\mu$s, and is illustrated in Fig.~\ref{Fig3}(a). 

The interaction described in Eqn.~\ref{eqn1} results in a dispersive shift ($\chi_m\hat{a}^{\dagger}_m\hat{a}_m\hat{c}^{\dagger}\hat{c}$) which leads to the qubit frequency being dependent on the photon number of each cavity, resulting in well-resolved transitions due to the high cooperativity~\cite{schuster2007resolving}. 
In the SNAP protocol, we use a combination of number selective qubit rotations ($\ket{g n_m} \leftrightarrow \ket{e n_m}$) and cavity displacements for universal control, as illustrated in Fig.~\ref{Fig3}(b). The pulse sequences are obtained through optimal control using the gradient ascent pulse engineering (GrAPE) algorithm ~\cite{khaneja2005optimal,heeres2017implementing}. The optimal control pulse used to prepare Fock state $\ket{1}$ in mode 2 is shown in the inset of Fig.~\ref{Fig3}(d), where we also show Wigner tomography of the state. 

Resolved transmon drives resonant with transitions corresponding to different photon numbers can also be used to blockade selected states and thus carve the allowed Hilbert space that is connected by a single cavity drive tone. This is illustrated in Fig.~\ref{Fig3}(e), where a resonant $\ket{g2_m}-\ket{e2_m}$ drive hybridizes the states, selectively shifting their energies by the Rabi drive strength $\Omega$ (for $\Omega \ll \chi_m$). The cavity mode thus inherits an anharmonicity $\Omega$, with the $\ket{2}$ state being blockaded~\cite{bretheau2015quantum}. A sufficiently weak cavity drive ($\epsilon\ll\Omega$) therefore results in a Rabi oscillation, as shown in Fig.~\ref{Fig3}(e), which can be used to prepare an arbitrary qubit state of $\ket{0},\ket{1}$ in any cavity mode. This scheme can be generalized to perform universal operations on qudits realized in any mode, and to prepare multimode entangled states with appropriate choice of the blockade drive~\cite{chakramhe2020}.

\begin{figure}[]
\includegraphics[width=0.5\textwidth]{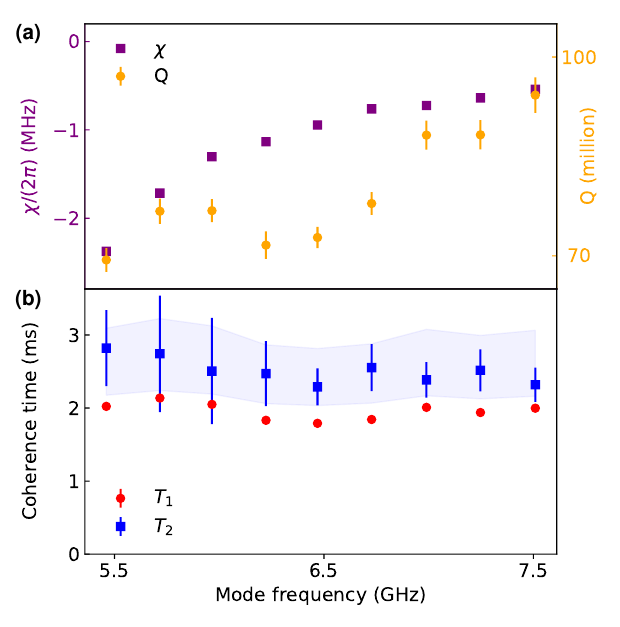}
\caption{\textbf{Dispersive shift and mode lifetimes}. (a) (left) Dispersive shift and (right) quality factors for the first 9 modes of the multimode storage cavity [MM2(5N5)]. (b) Cavity mode decoherence times ($T_1, T_2$) for the first 9 modes. The blue region represents the theoretical mode $T_2$ limits arising from additional dephasing due to the thermal population of the transmon ($n_{\mathrm{th}}^q = 1.2\pm0.5\%$).}
\label{Fig4}
\end{figure}

The fidelity of these cavity control protocols is limited primarily by decoherence arising from the transmon during the gate operation. The gate speed is set by the dispersive shift, shown as a function of the mode frequency for MM2(5N5) in Fig.~\ref{Fig4}(a, left). In the case of the SNAP gate and resonant sideband SWAP operations, the photon is present in the transmon during the gate, with the minimum transmon-induced infidelty scaling as $~1/(\chi T_q)$ up to prefactors, where $T_q$ is the minimum of the qubit decay and decoherence time. In photon blockade, although the transmon is never directly occupied, optimizing the drive strength to minimize leakage and blockade induced Purcell decay still results in the same minimum infidelity. In MM2(5N5), the $\ket{f0}-\ket{g1}$ SWAP is additionally affected by dephasing induced by driving through the readout port, with the large drive powers resulting in off-resonant population of the readout cavity (see SI). This effect can be mitigated by driving the storage cavity or the transmon directly. The gate fidelities are also a function of the instrinsic quality factors of the modes, which range from $65-95\times10^6$, as shown in Fig.~\ref{Fig4}(a, right). This results in an expected additional infidelity of $~0.1~\%$ for the sideband and SNAP gates, and $\sim 1-2~\%$ for the longer blockade gates.

We characterize the decay and decoherence times of the cavity modes by $T_1$ and Ramsey measurements, the results of which are shown in Fig.~\ref{Fig4}(b). Here, each cavity mode is initialized in the $\ket{1}$ and $\ket{0} + \ket{1}$ states using cavity Rabi oscillations performed in the presence of photon blockade, as shown in Fig.~\ref{Fig3}(d). The $T_1$ times of all the cavity modes were $\sim2$ ms, while the $T_{2}$ times range from $2-3$ ms. The deviation of $T_2$ from $2T_1$ is consistent with additional dephasing from cavity frequency fluctuations arising from thermal excitations of the transmon (blue band in Fig.~\ref{Fig4}(b)). These coherence times are nearly two orders of magnitude better than those that have been reported in any multimode cQED system. The coherence times of any of these cavity modes is also comparable to the longest reported in single or few-mode 3D cQED systems.

In summary, we have demonstrated a new flute method for creating high quality factor seamless cavities with tailored mode-dispersion and mode-qubit couplings, ideally suited for creating multimode circuit-QED systems with high cooperativities across all modes. As quantum systems increase in volume and processor size, one of the most important challenges is the hardware overhead of lines and attendant equipment required for the control of every qubit or cavity mode. In this work we have demonstrated---with a single control line---a variety of schemes for universal control of $\sim$10 cavity modes using the nonlinearity of a single transmon. This is an important step for realizing cavity-based random access memories and processors, and for engineering photonic platforms ideally suited for exploring quantum many-body physics with microwave photons utilizing the toolbox of quantum optics. In principle, this control can be extended to $\sim$1000 cavity modes by leveraging state-of-the-art accelerator cavity technology to improve cavity coherence times.  While we have demonstrated quantum control of a single multimode cavity, these systems can also act as modules which can be coherently coupled~\cite{leung2019deterministic} to build larger processors and perform quantum error correction with minimal hardware overhead.

\begin{acknowledgments} 
We thank Gerwin Koolstra for useful discussions and experimental assistance. This work was supported by the the Samsung Advanced Institute of Technology Global Research Partnership and the ARO grant W911NF-15-1-0397. This work is funded in part by EPiQC, an NSF Expedition in Computing, under grant CCF-1730449. We acknowledge the support provided by the Heising-Simons Foundation. D.I.S. acknowledges support from
the David and Lucile Packard Foundation. This work was partially supported by the University of Chicago Materials Research Science and Engineering Center, which is funded by the National Science Foundation under award number DMR-1420709.  Devices were fabricated in the Pritzker Nanofabrication Facility at the University of Chicago, which receives support from Soft and Hybrid Nanotechnology Experimental (SHyNE) Resource (NSF ECCS-2025633), a node of the National Science Foundation’s National Nanotechnology Coordinated Infrastructure.
\end{acknowledgments} 
\bibliography{thebibliography}
\end{document}


\title{Seamless high-Q microwave cavities for multimode circuit QED: Supplementary Information} 
\date{\today}
\maketitle

\section{Fabrication of flute cavities}
\begin{figure*}
   \centering
    \includegraphics[width= 0.85\textwidth]{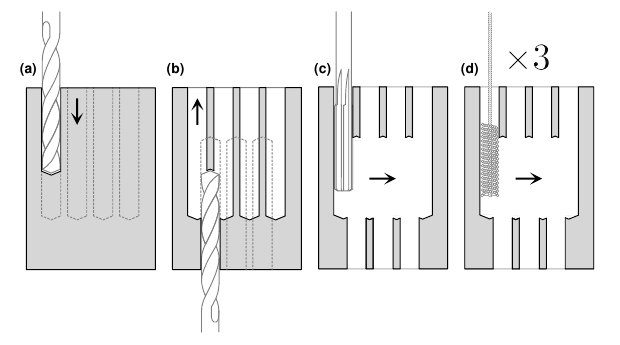}

    \caption{Schematic illustration of the creation of a rectangular cavity using the flute method. (a) First, undersized holes are drilled from the top of the stock using a Jobber drill. (b) the stock is flipped and the steps in (a) are repeated with the drill moved over half of a hole spacing. (c) Reaming is done to ensure the holes are perfectly straight and to create a more uniform machined surface. This step is repeated for the top and bottom holes. (d) To further refine the surface finish and remove burrs, honing is done using a ball hone. Three passes are done with successively finer grit to both top and bottom holes, creating an even polished surface in preparation for etching.}
  	\label{flute}
\end{figure*}

The flute technique creates low-loss seamless cavities by using intersecting holes to create a cavity volume, with the diameter setting the waveguide cutoff frequency of the hole. The choice of cutoff frequency determines the propagation constant $\beta$ of the cavity field through the hole. For the $\mathrm{TE_{nm}}$ mode, the cutoff frequency is:
\begin{equation}
    f_{c_{nm}}=\dfrac{k_{c}}{2\pi\sqrt{\mu\epsilon}},
\end{equation}
where $k_c=p'_{nm}/a$ is the cutoff wavenumber, $p'_{nm}$  is the $\mathrm{m^{th}}$ root of the derivative of the $\mathrm{n^{th}}$ bessel function of the first kind ($J'_n$), and $a$ is the hole radius. For a frequency with wavenumber $k$, the propagation constant through the waveguide is $\beta_{nm}=\sqrt{k^2-k_{c}^{2}}$ \cite{pozar2009microwave}, which becomes imaginary when the propagating frequency is lower than the cutoff frequency of the fundamental waveguide mode. By making the hole diameter small, the cutoff frequency is made much higher than that of the cavity modes ($f_c\gg f_i$). This leads to an exponential attenuation of the cavity field along the length of the hole ($L$), resulting in the quality factor limit from evanescent coupling to the vacuum scaling as $Q_{\mathrm{ext}}\propto e^{-\beta_{nm}L}$. We achieved $Q_{\mathrm{ext}} \gg 10^9$ by choosing $L > a$, thereby completely mitigating the effect of evanescent loss.

We create seamless cavities from monolithic blocks of high purity ($>99.999\%$) aluminum. A schematic of this process is illustrated in SFig.~\ref{flute}. The process---from left to right---is as follows: in SFig.~\ref{flute} (a) we begin by drilling holes undersized from the desired diameter ($d=4.76$ mm) by $\sim50-100~\mu$m along the top of the stock at our desired spacing ($l_{s}=6.03$ mm). We use a standard uncoated Jobber style drill bit with parabolic flutes for reduced galling and wear. After the top holes are drilled to the desired depth, the stock is flipped and the process is repeated with the bottom holes displaced by half the center-center hole distance from the top holes, as shown in SFig.~\ref{flute} (b). The overlap of these holes form the cavity volume, depicted in white. An additional reaming step is performed to bring the holes to their final dimension, and to ensure surface uniformity and hole straightness, as shown in SFig.~\ref{flute} (c). This is repeated for both top and bottom holes. Finally a ball hone, made of silicon carbide abrasives, is used to create a uniformly smooth surface and remove internal burrs that may form during drilling and honing, as shown in SFig.~\ref{flute} (d). This process is repeated three times using hones made of successively finer abrasive media. The ductility and galling properties of the aluminum causes the pores of the abrasive to "fill", leading to a drop in honing efficacy after many successive holes. It is thus recommended to inspect the surface finish regularly at this point, replacing the hones as necessary, and to use a lubricant to reduce heat caused by friction. This is repeated for the top and bottom holes. 

\begin{figure*}[t]
  \begin{center}
    \includegraphics[width=1.0 \textwidth]{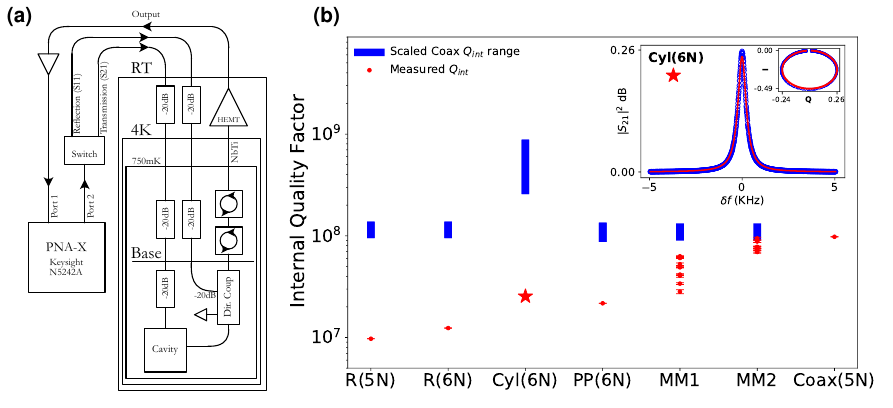}
    \caption{(a) A depiction of the setup for the measurement of cavity quality factors without a transmon. $S_{21}$ and $S_{11}$ measurements were done concurrently by using a directional coupler on the output line of the cavity. (b) A comparison of quality factors for the various flute geometries. We show the measured quality factors (red circles), and the expected theoretical quality factor range (blue bars), after scaling the coaxial cavity [Coax(5N)] quality factor by the respective $S_{e}$ and $S_{m}$ values for each cavity. (inset) A plot showing the cylindrical flute cavity [Cyl(6N)] resonator spectroscopy in transmission ($S_{21}$).}
  	\label{cavity_characterization}
  \end{center}
\end{figure*}

After the holes are created, the cavities undergo an etching process. The process removes $\sim100~\mathrm{\mu m}$ of material from the surface, reducing surface damage caused by the machining process. Etching was done using Transene Aluminum Etch A at $50^{\circ}\mathrm{C}$ with constant agitation from a magnetic stirrer. All cavities underwent etching for 4 hours total. MM1(5N5), R(5N), P(6N), R(6N), Cyl(6N) and Coax(5N) all underwent two 2 hour etches with the etchant replaced at each interval. The MM2(5N5) cavity was etched over the same time span with more frequent changes of the etchant, with an initial 2 hour etch followed by four 30 minute etching steps. The more frequent exchanges of etchant were done to compensate for the larger surface area of the cavity and to ensure a more even etch rate throughout the process. The data for MM2(5N5), R(6N) and R(5N) presented in SFig.~\ref{cavity_characterization}(b) was taken after the cavities were re-etched for an additional 4 hours using their respective recipes.

\section{Characterization of various flute cavity geometries}

\begin{figure}
    \includegraphics[width=.45 \textwidth]{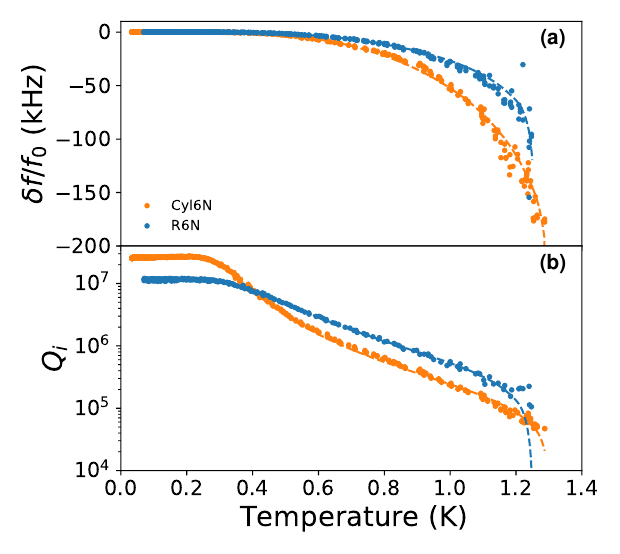}
    \caption{ (a) The change in the $\mathrm{TM_{011}}$ mode frequency of the cylindrical flute cavity [C(6N)], and the  $\mathrm{TE_{101}}$ mode of the rectangular pan-pipe cavity [R(6N)], as a function of temperature. (b) The corresponding internal quality factors as a function of temperature. The fits (dashed lines) give  $T_c\sim1.31\mathrm{K}$ and $1.25\mathrm{K}$ for the C(6N) and R(6N) cavity, respectively. From the frequency fits, we can extract the London penetration lengths for both cavities, giving values of $235\pm3$~nm and $37.6\pm0.9$~nm, respectively. 
    }
    \label{cyl_QFvsT}
\end{figure}

Cavity spectroscopy was performed in both reflection and transmission to take advantage of the better SNR of transmission and allow for the independent measurement of the coupling Q of each port. All measurements without a qubit were taken in an Oxford Triton dry-dilution refrigerator with a mixing chamber (MXC) temperature of $\sim36$ mK. A schematic of the fridge wiring, filtering and the RT microwave components is depicted in Fig.~\ref{cavity_characterization} (a). To determine quality factors, both reflection and transmission measurements were done concurrently by switching the output of a network analyzer to the appropriate port. The reflection measurements ($S_{11}$) were taken by passing the signal through the $-20$ dB port of a Marki C20-0116B, 1-16 GHz directional coupler, which was preceded by two $-20$ dB attenuators anchored at MXC and 4K plate. For the transmission ($S_{21}$) input line, an additional $-20$ dB was added at the MXC plate. The output of the directional coupler was followed by a pair of Quinstar OXE89, 4-12 GHz isolators, before passing through an LNF 2-12 GHz HEMT amplifier located at the $4$K plate of the DR. Additional amplification was done via a Miteq 3-12 GHz amplifier at RT. CW measurements were taken using a Keysight N3242A PNA-X. All measurements were done in a high power regime, with average photon numbers $\bar{n}>1000$.

A total of 7 cavities made using the flute method were measured, the results of which are listed in Fig.~1 (e) of the main text. Of these cavities, R(5N), R(6N), and Cyl(6N) were measured via spectroscopy, and MM1(5N5) was characterized by ringdown measurements. MM2(5N5) and P(6N) were characterized using a coupled qubit using methods outlined in section~\ref{mode_coherence_with_qubit}. An overview of all the measured coherences can be found in SFig.~\ref{cavity_characterization} (b) (red points). This figure also shows the theoretical range of quality factors (blue bars) if the coaxial cavity [Coax(5N)] quality factors are scaled according to $S_e$ and $S_m$ of each cavity geometry, assuming that the dominant loss is one or the other. In reality, the total internal quality factor is determined by contributions of both magnetic and dielectric loss mechanisms. Also shown is the measured $\mathrm{TM_{011}}$ mode (denoted by the red star) of the cylindrical cavity (inset) with magnitude and I-Q data and fits plotted (red line). The cylindrical cavity was made out of 6N aluminum and underwent the same manufacturing and etching steps as the other cavities. The transmission and reflection data for the cavities were fit using the following equations respectively:

\begin{equation}
    |S_{21}|^{2}=\bigg{|}\dfrac{2\sqrt{\tilde{\kappa}_{1}\tilde{\kappa}_{2}}}{(\omega-\omega_{0})+i(\tilde{\kappa}_{1}+\tilde{\kappa}_{2}+\kappa_{i})}\bigg{|}^{2},
\end{equation}

\begin{equation}
    |S_{11}|^{2}=\bigg{|}\dfrac{i(\omega-\omega_{0})+(\kappa_{i}-\tilde{\kappa}_{1})}{i(\omega-\omega_{0})+(\kappa_{i}+\tilde{\kappa}_{1})}\bigg{|}^{2},
\end{equation}
where $\tilde{\kappa}_{n}=\kappa_{n}+i\gamma_{n}$ is the rate of coupler $n$ with the addition of a complex loss $\gamma_{n}$. Under normal circumstances, $\kappa_{n}\gg\gamma_{n}$, giving a symmetric line shape. 

Of the cavities measured, the quality factor of the cylindrical flute deviated the most, having the lowest numerically determined magnetic surface participation ratio for its $\mathrm{TM_{011}}$ mode, leading to a measured internal quality factor ($Q_{\mathrm{int}}$) over an order of magnitude lower than predicted, assuming the same superconducting properties of the Coax(5N) cavity. SFig.~\ref{cyl_QFvsT} shows the temperature dependence of both the internal quality factor and mode frequency as a function of temperature for the fundamental modes of both the rectangular \textit{pan-pipe} [R(6N)] and cylindrical cavity [Cyl(6N)]. Both cavities were made from the same 6N aluminum stock, underwent the same manufacturing steps shown in SFig.~\ref{flute}, were treated using the same etching steps outlined in section I above, and were measured in the same cooldown using separate but identical measurement chains as outlined in SFig.~\ref{cavity_characterization} (a). The large shift in frequency of the cylindrical cavity from $T=36$ mK to $T_c$ indicates a high kinetic inductance fraction and penetration depth. To determine the London penetration depth and ratio of scattering to coherence length, the data of SFig.~\ref{cyl_QFvsT} was fit using:

\begin{equation}
    \dfrac{\delta f(T)}{f_o}=p_{\mathrm{mag}}\bigg{(}\dfrac{\delta\sigma_2(\omega, T)}{\sigma_2(\omega, 0)}\bigg{)}^{\nu} 
\end{equation}
where $\sigma_2(T)$ is the imaginary component of the BCS conductivity and $\nu$ is a scaling parameter based on the ratios of the mean-free path ($l$) and the coherence length ($\xi_o$). For aluminum, the mean free path is typically much shorter than the coherence length ($\xi_o \gg l$), therefore $\nu=-1/3$ (the "dirty" or Pippard limit) \cite{Nam1967BCS}. The parameters $p_{\mathrm{mag}}$ and $T_c$ are determined from fitting the data, where $p_{\mathrm{mag}}=\lambda_L  S_m$. The fits for the internal quality factor as a function of temperature takes the form:

\begin{equation}
    Q_{int}(T)=\bigg{(}\dfrac{1}{Q_{int,\mathrm{max}}}+\dfrac{p_{\mathrm{mag}}}{Q_{s}(T)}\bigg{)}^{-1}
\end{equation}
where $Q_{s}(T)=X_s(T)/R_s(T)$, which reduces to $Q_{s}(T)\sim\sigma_2(\omega, T, T_c)/\sigma_1(\omega, T, T_c)$ \cite{ZmuidzinasSupercond}. $Q_{s}$ is the quality factor contribution limited by magnetic defects of the superconductor. As such, the contribution scales with the magnetic participation of the field.  

Fitting the temperature dependant data presented in SFig.~\ref{cyl_QFvsT} for frequency and internal quality factor shows good agreement with BCS theory, with the BCS conductivities numerically computed for every temperature. From this, we can readily calculate the London penetration depth $\lambda_{\mathrm{L}}$ from the $\mathrm{p_{mag}}$ and the numerically determined participation ratios from HFSS simulations. For the cylindrical cavity, $S_m = 187~\mathrm{m^{-1}}$, and $p_{\mathrm{mag}}=4.61\times10^{-5}$, giving a $\lambda_{\mathrm{L}}=235\pm 3~$nm. In the Pippard limit, the London penetration length scales as $\lambda_{\mathrm{L}}\approx\lambda_o(1+\xi_o/l)^{1/2}$, where $\lambda_o=16~$nm and $\xi_o\sim1600~$nm are the intrinsic London and coherence lengths for aluminum, respectively \cite{tinkham2004introduction}. This indicates that the mean-free path is much lower than the coherence length, meaning that the superconductor is extremely ``dirty'', leading to more of the supercurrent interacting with magnetic defects of the material. In contrast, doing the same analysis as above for the R(6N) flute cavity gives us a London penetration depth of $37.6\pm0.9$ nm, given a numerically simulated participation ratio of $S_m=505~\mathrm{m^{-1}}$. This value of $\lambda_{\mathrm{L}}$ is nearly an order of magnitude lower than that of the cylindrical cavity, showing that the surface defects are much better mitigated in the 6N rectangular flute design.

While the maximum quality factor ($Q_{\mathrm{int,max}}$ in SEqn. 4) at base temperature---where thermal quasiparticle formation should be zero---is still higher for the cylindrical cavity, the additional field penetration means that the cylindrical cavity is still more susceptible to loss due to non-equilibrium quasiparticle formation. An additional contribution to $Q_{\mathrm{int,max}}$ is electric field participation with lossy surface oxides. The $\mathrm{TM_{011}}$ mode of the cylindrical cavity has a participation ratio of $S_e=64~\mathrm{m^{-1}}$ compared to an $S_e$ of $409~\mathrm{m^{-1}}$  for the rectangular design. This disparity likely explains the differences in $Q_{\mathrm{int}}$ at $T \ll T_c$ for the R(6N) cavity.

The implications of the above results highlight an important consideration when designing flute cavities. While the overlap of holes creates the cavity volume in both cases, in the rectangular cavity the surface created by the drilling operation, and the subsequent honing steps used to smooth that surface, directly participate in the field. The cylindrical cavity's participating surfaces are instead left untouched by the subsequent smoothing and refining steps, leaving the participating surface with more defects. These defects lead to a larger surface area which can affect the etching process. In niobium cavities surface roughness has been shown to negatively impact etching efficacy and quality factor, however in aluminum coaxial cavities the connection between surface roughness and performance after etching is not as strongly correlated [S5, S6]. An alternative approach to solve this problem for the cylindrical design is one where the overlapping holes are axial instead of radial in a \textit{wood piling} configuration, allowing for the direct polishing of the participating surfaces during the refining steps.  

\section{Fabrication of the transmon qubit}
\label{appendix:b}
The transmon qubit was fabricated on a $430~ \mu$m thick C-plane (0001) Sapphire wafer with a diameter of $50.8~$mm. The wafer was cleaned with organic solvents (Toluene, Acetone, Methanol, Isopropanol, and DI water) in an ultrasonic bath to remove contamination, then annealed at $1200~^{\circ}\mathrm{C}$ for 1.5 hours. Prior to film deposition, the wafer underwent a second clean with organic solvents (Toluene, Acetone, Methanol, Isopropanol, and DI water) in an ultrasonic bath. The junction was made out of aluminum using a combination of optical and ebeam lithography. The base layer of the device, which includes the capacitor pads for the transmon, consists of $120$ nm of Al deposited via electron-beam evaporation at $1 A^{\circ} /s$. The features were defined via optical lithography using AZ MiR 703 photoresist and exposure by a Heidleberg MLA150 Direct Writer. The resist was developed for 1 minute in AZ MIF 300 1:1. The features were etched in a Plasma-Therm inductively coupled plasma (ICP) etcher using chlorine based etch chemistry (30 sccm $\mathrm{Cl}_2$, 30 sccm $\mathrm{BCl}_2$, 10 sccm Ar). This was followed by a second layer of optical patterning and subsequent thermal evaporation of 50 nm of Au, for the allignemt marks used for ebeam lithography. The resist was subsequently removed by leaving the wafer in $80^{\circ}$C N-Methyl-2-pyrrolidone (NMP) for 4 hours. The junction mask was defined through electron-beam lithography of a bi-layer resist (MMA-PMMA) in the Manhattan pattern using a Raith EBPG5000 Plus E-Beam Writer, with overlap pads that allow for direct galvanic contact to the optically defined capacitors. The resist stack was developed for 1.5 minutes in a solution of 3 parts IPA and 1 part DI water. Before deposition, the overlap regions on the pre-deposited capacitors were milled \textit{in-situ} with an argon ion mill to remove the native oxide. The junction was then deposited with a three step electron-beam evaporation and oxidation process. First, an initial $35$ nm layer of Al was deposited at $1$ nm/s at an angle of $29^\circ$ relative to the normal of the substrate, azimuthally parallel to one of the fingers in the Manhattan pattern. Next, the junction was exposed to $20$ mBar of a high-purity mixture of $\mathrm{Ar}$ and $\mathrm{{O}_2}$ (80:20 ratio) for 12 minutes to allow the first layer to grow a native oxide. Finally, a second $120$ nm layer of Al was deposited at $1$ nm/s at the same angle relative to the normal of the substrate, but azimuthally orthogonal to the first layer of Al. After evaporation, the remaining resist was removed via liftoff in $80^{\circ}$C NMP for 3 hours, leaving only the junction directly connected to the base layer. After both the evaporation and liftoff, the device was exposed to an ion-producing fan for 30 minutes to avoid electrostatic discharge of the junction.

\section{Improving transmon coherence and thermal population}

\begin{figure*}[t]
  \begin{center}
    \includegraphics[width=1.0 \textwidth]{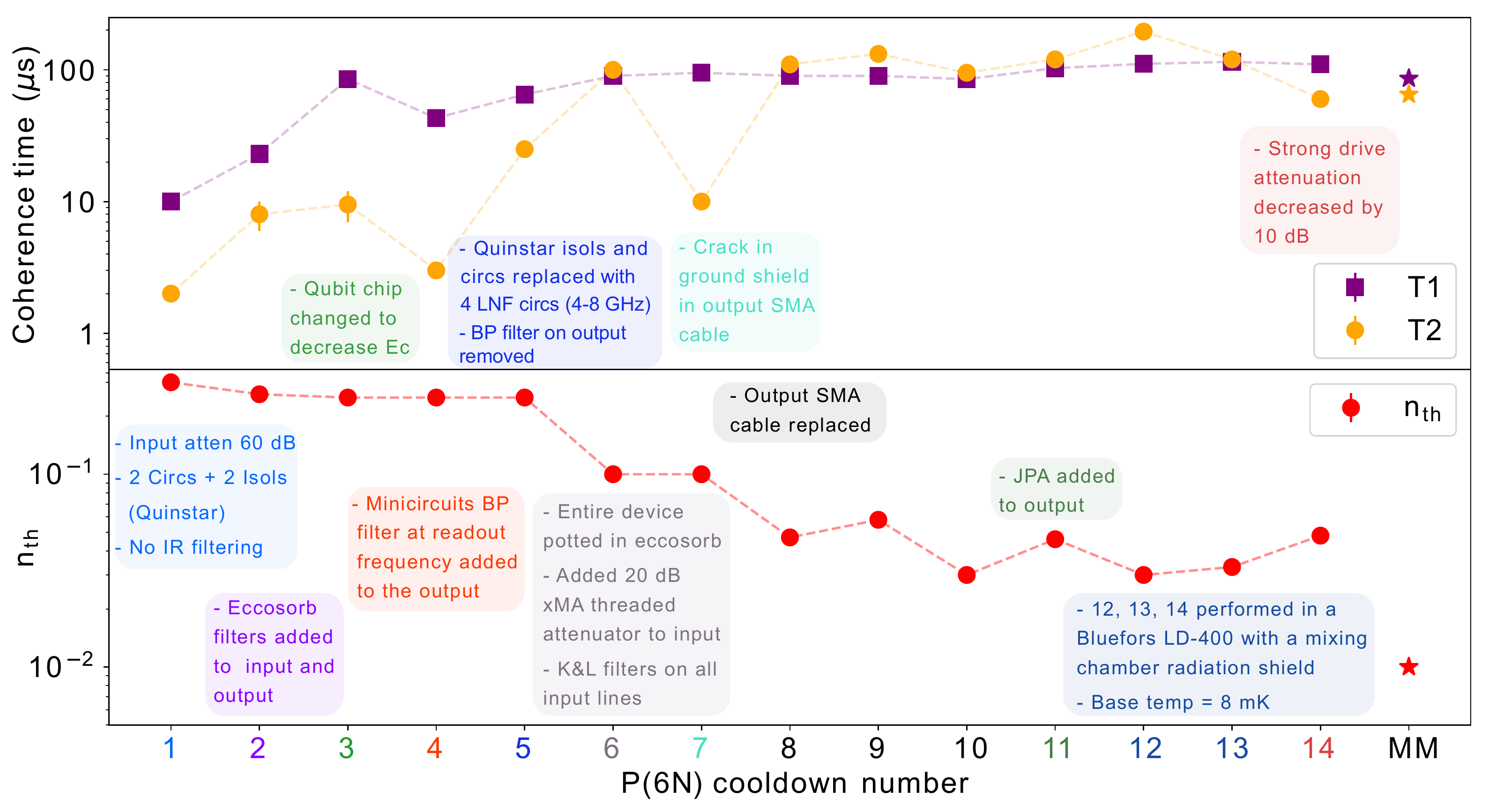}
    \caption{Evolution of (top) qubit coherence and (bottom) thermal population ($n_{\mathrm{th}}^q$) for P(6N) over the course of 14 cooldowns. The primary changes that were made in each cooldown are listed in the legend (color coded with the cooldown number). The most important jumps to be noted are the reduction the qubit temperature from potting the device in IR absorber (6) and the improvement in the transmon $T_2$ in cooldown (5) from changes to the output line. These results were applied to the multimode device [MM2(5N5)], providing the coherence and temperature indicated by the $\star$.}
  	\label{qubit_coherence_improvements}
  \end{center}
\end{figure*}

We improved the transmon coherence and the thermal population by changing various aspects of the sample shielding and the filtering of the lines over repeated cooldowns of the same device [P(6N)].  The coherence times of the transmon were characterized through standard Ramsey and $T_1$ measurements. The qubit temperature was measured using the contrast of the Rabi oscillation on the $\ket{e}-\ket{f}$ transition with and without a $\pi$-pulse on the $\ket{g}-\ket{e}$ transition. The temperature of the readout resonator was inferred from the dephasing of the transmon. 

A summary of the changes made in each cooldown and the effect on the transmon coherence and the qubit temperature is summarized in SFig.~\ref{qubit_coherence_improvements}. The qubit $T_1$ was in the 50-100 $\mu$s range with the early iterations of the filtering. Repeated cooldowns were necessary to reduce thermal populations of the qubit and the readout resonator. The thermal population of the transmon was initially high because of inadequate shielding of IR radiation at frequencies above the superconducting gap, which likely lead to the creation of quasiparticles. This radiation was leaking in from the higher temperature stages of the fridge due to the sample package being inadequately light tight. This was consistent with a 3-fold reduction in the thermal population after potting in the entire device in IR absorber (ECCOSORB). Unlike 2D devices with a large ground plane, quasiparticles generated in 3D transmons likely lead to a transmon heating event due to a lack of well-thermalized ground plane.

The $T_2$ of the transmon was limited by the thermal population of the readout resonator.  This was likely due to photons at the readout frequency leaking in through the microwave lines. Thermal photons coming through the input line from the 4K stage of the dilution fridge were reduced by better thermalization of the attenuators at the base stage. This was done by using a thin film cryo-attenuator, described in~\cite{Yeh2017}, and threaded copper cryo-attenuators from XMA. Thermal readout photons were found to primarily be coming in from the output line, coupled to the readout cavity through an overcoupled port with $Q_c\sim15000$. This was improved by using circulators that could be better thermalized at the base stage (Low Noise Factory), particularly the first circulator in the output chain, closest to the readout resonator. The thermal population of the readout was further reduced by the addition of a weak ECCOSORB filter on the output line, inside the sample can, and as close to the output coupler as allowed by geometry. The final wiring and shielding configuration incorporating these improvements for measurements on device P(6N) is shown in SFig.~\ref{wiring diagram} (a).

These experiments led to the following best practices for coupling the transmon to the multimode cavity [MM2(5N5)]: the transmon chip is inserted into the multimode cavity through the readout cavity and held using a two-piece copper mount, with a flat shallow channel in the bottom piece for the transmon chip. The qubit mount halves are clamped together, holding the qubit chip, and sealed using an indium wire compressed between the two halves to create a light-tight interface. An indium wire is also used to seal Al plates covering the top and bottom flute holes. The device is heat sunk to a plate that is connected to the lid of the box that contains the sample, both made out of OFHC Copper. The sample is surrounded by a can containing two layers of $\mu$-metal shielding, with the inside of the inner layer connected to a can made out of copper shim that is attached to the OFHC copper can lid and painted on the inside with Berkeley black. The can attaches to the base stage of a Bluefors LD-400 dilution refrigerator (7.5-8 mK). The base stage of the refrigerator had an additional radiation shield, which lead to fewer IR photons at the device. 
With this protocol, we achieve transmon coherence times $T_1,T_2 \sim 80-100,60-140 \mu$s, and thermal occupations of $~1\%$, which is comparable to state-of-the-art values in 3D cQED systems [S5-S7].

\section{Control instrumentation}
\begin{figure*}[t]
  \begin{center}
    \includegraphics[width= \textwidth]{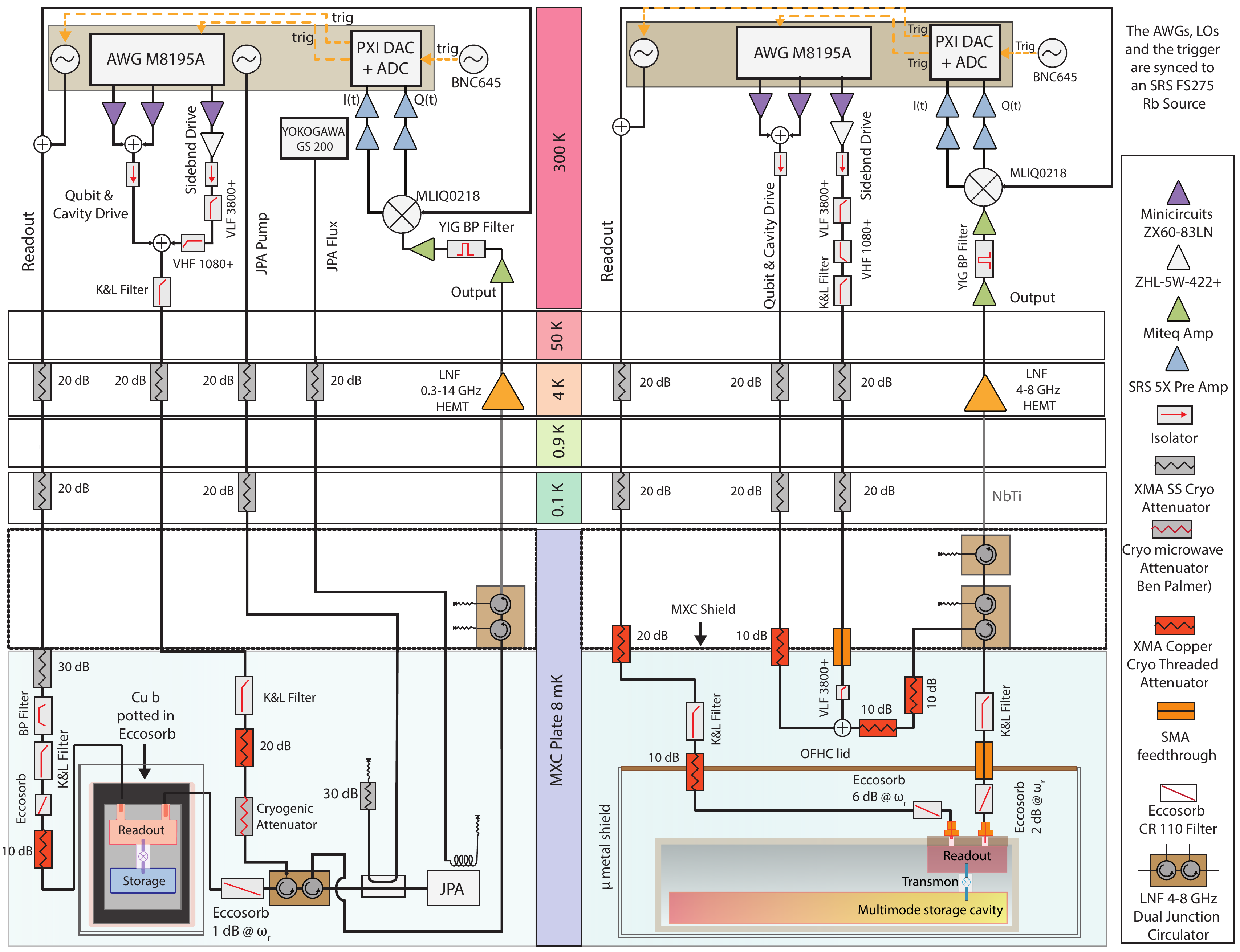}
    \caption{Cryogenic setup and wiring diagram for (left) cooldown 12 the P(6N) device,  and (right) the multimode device MM2(5N5). The coherences and thermal populations that result are shown in SFig.~\ref{qubit_coherence_improvements}. While P(6N) was made light tight by potting in IR absorber, MM2(5N5) was made light tight using indium seals.}
  	\label{wiring diagram}
  \end{center}
\end{figure*} 
 A schematic of the cryogenic setup, control instrumentation, and device wiring is shown in SFig.~\ref{wiring diagram}. 
 All controls and readout are performed through the readout cavity, by driving at the qubit and storage mode frequencies. The pulses are directly synthesized using a 4-channel, 64 GSa/s arbitrary waveform generator (Keysight M8195A). The combined signals are sent to the device after being attenuated at each of the thermal stages, as shown in SFig.~\ref{wiring diagram}. The transmitted output signal from the readout resonator passes through three cryogenic circulators (thermalized at the base stage) and amplified using a HEMT amplifier (anchored at 4K). Outside the fridge, the signal is filtered (tunable narrow band YIG filter with a bandwidth of 80 MHz) and further amplified. The amplitude and phase of the resonator transmission signal are obtained through a homodyne measurement, with the transmitted signal demodulated using an IQ mixer and a local oscillator at the readout resonator frequency. The homodyne signal is amplified (SRS preamplifier) and recorded using a fast ADC card (Keysight M3102A PXIe 500 MSa/s digitizer).

\section{Multimode system Hamiltonian}
A transmon coupled to a set of cavity modes is described by the Hamiltonian,
\begin{equation}
\begin{split}
	\hat{H} = \nu_\mathrm{t} \hat{c}^\dagger \hat{c} &+ E_J\left(\cos\left(\phi\right)+\frac{\phi^{2}}{2} - 1\right)\\
	&+\sum_{k=1}^n \left[ \nu_k\hat{b}_k^\dagger \hat{b}_k +  g_k(\hat{b}_k + \hat{b}^\dagger_k )(\hat{c} + \hat{c}^{\dagger}) \right] .
\end{split}
\end{equation}
Here $b^{\dagger}_{k},b_{k}$ and and $c^{\dagger},c$  are the bare operators for creation and annihilation of a photon in cavity mode $k$, and the transmon, respectively. The frequencies are $\nu_k$ and $\nu_t$, with $g_k$ representing the dipole interaction between them. We note that we will identify one of the modes as the readout mode ($r$) and represent the modes of the multimode cavity as $m$. We diagonalize the linear part of the coupled system, and rewrite the Hamiltonian in terms of the dressed transmon-cavity variables ($\tilde{c},\tilde{b}_k$),
\begin{equation}  \label{H_total}
	\hat{H} =  \sum_{k=1}^n \left[ \tilde{\nu}_k\tilde{b}_k^\dagger \tilde{b}_k\right] + \tilde{\nu}_\mathrm{t} \tilde{c}^\dagger \tilde{c} + E_J\left(\frac{\phi^4}{4!} - \frac{\phi^6}{6!}  + \ldots\right),
\end{equation}
and rewrite the junction phase in terms of the dressed operators:
\begin{equation}
\phi =  \phi_c\left\{\beta_{c}\left(\tilde{c}+\tilde{c}^{\dagger}\right) + \sum_{k}\beta_{k}\left(\tilde{b}_{k} + \tilde{b}^\dagger_{k}\right)\right\}.
\end{equation}
In the above, $\phi_{t} = \left(2 E_C/E_J\right)^{\frac{1}{4}}$ is the zero-point fluctuation of the phase of the junction, and $\beta_{k} \sim (g_k/\Delta_k)$ is the participation of each mode in the junction. Enforcing the commutation relations between the dressed operators implies that $\beta^{2}_{t} + \sum_{k}\beta_{k}^{2} = 1$, where $\beta_t$ is the participation of the transmon in junction phase. This approach is similar to that used in black-box quantization~\cite{nigg2012black}.  Here, we use the admittance of the system as seen by the junction port---extracted using driven modal simulations performed using ANSYS HFSS---to obtain the impedance of each mode. The zero point phase fluctuation of each mode is $\tilde{\phi}_k = \sqrt{2\pi G_{Q}Z_k}$, where $G_{Q} = 2e^2/h$ is the quantum of conductance, and $Z_{k}= \sqrt{L_k/C_k}$ is the impedance of the mode. We infer the participation factors from the impedance using $\beta_k/\beta_t = \sqrt{Z_k/Z_t}$. These quantities can be used to infer the coupling strengths, the self-Kerr and cross-Kerr interactions of the transmon and the cavity modes. For the single-mode cavity, the $\alpha$ and the $g$ extracted from simulation match the experimentally measured values to within $5\%$. For the multimode cavity, the black-box quantization was found to underestimate the $\alpha$ by $10-20\%$. In this case, the $\alpha$ was inferred from $E_c\approx\alpha$, extracted directly from the capacitances obtained from ANSYS Q3D. We note that in order to get coupling strengths ranging from  $50-150$ MHz in the multimode cavity, we require transmon capacitor pad lengths of $\sim5$ mm. In order to have a sufficiently large $E_c \sim 140$ MHz, our transmon design had asymmetric pads, with the $E_c$ being controlled by the size of the smaller pad. 

\section{Cavity control using the Josephson non-linearity}

The Josephson non-linearity of the transmon can be used to perform quantum operations on any of the cavity modes through their participation in the junction phase. In this work, we perform these operations by only driving the readout cavity at different frequencies. The coupling between the readout cavity and the transmon results in an indirect transmon drive, as can be seen by performing a dispersive transformation to lowest order:
\bea
 \hat{H}_d = \epsilon(t)\left(\hat{b}_r + \hat{b}_r^{\dagger}\right) &=& \epsilon(t)\left(\tilde{b}_r e^{-i\tilde{\omega}_r t}+\beta_{r}\tilde{c}e^{-i\tilde{\omega}_c t} + \mathrm{c.c.}\right),\nonumber\\
 &=& \frac{\tilde{\epsilon}}{2}\left(\tilde{b}_r e^{-i\tilde{\delta}_r t} +  \beta_{r}\tilde{c}e^{-i\tilde{\delta}_t t} + \mathrm{c.c} \right).
 \label{Hd}
\eea 
In the expression above, where $\epsilon(t)  =\tilde{\epsilon}\cos(\omega_d t)$, we have used the transformation $\hat{U}_{RF} = e^{-i\left(\tilde{\omega}_{t}\tilde{c}^{\dagger}\tilde{c} + \sum_{k}\tilde{\omega}_k\tilde{b}^{\dagger}_{k}\tilde{b}_{k}\right)t}$ to move to a frame rotating at the natural frequencies of the dressed transmon and the cavity modes, and dropped all counter rotating terms. $\delta_{r/t} = \omega_{r/t} -\omega_d$ are the detunings of the drive from the readout and transmon frequencies. To account for both of the drive terms we perform a displacement transformation $ 
\mathcal{D} = e^{\left(\xi(t)\tilde{b}_r^{\dagger} +\eta(t)\tilde{c}^{\dagger}-\mathrm{c.c.}\right)}$, resulting in:
\be
\tilde{b_r}\rightarrow\tilde{b}_r + \xi(t), \tilde{c}\rightarrow \tilde{c} + \eta(t), \hat{H} \rightarrow \mathcal{D}^{\dagger} \hat{H}\mathcal{D} - i  \mathcal{D}^{\dagger}\dot{\mathcal{D}}.
\label{disptrasnform}
\ee
Choosing  $\xi,\eta$ to satisfy the classical equations of motion:
\be 
i\dot{\tilde{\xi}} +\delta_r\tilde{\xi} = \frac{\tilde{\epsilon}}{2},\quad i\dot{\tilde{\eta}} +\delta_c\tilde{\eta}_i = \frac{\tilde{\epsilon}}{2}\beta_r,
\ee
with $\xi = \tilde{\xi} e^{-i\delta_r t}$ and $\eta = \tilde{\eta} e^{-i\delta_c t}$, leading to the drive term in Eqn.~\ref{Hd} being cancelled by the Berry phase term in Eqn.~\ref{disptrasnform}.
In steady state this results in $\tilde{\xi} = \frac{\tilde{\epsilon}}{2\delta_r}$ and $\tilde{\eta} = \beta_r\frac{\tilde{\epsilon}}{2\delta_c}$.
\subsection{Charge-sideband transitions}
The charge-sideband transition rates are evaluated by expanding the Josephson non-linearity to quartic order after performing the transformations previously described. This gives:
  \begin{equation}
\begin{split}
H_I = \frac{E_c}{12}&\Big[\beta_t\tilde{c}e^{-i\tilde{\omega}_t t} + \beta_r\tilde{a}_re^{-i\tilde{\omega}_r t} \\
&+ \sum_m\beta_m\tilde{a}_m e^{-i\tilde{\omega}_m t}
+ \beta_r\xi_d e^{-i\tilde{\omega}_d t} +  \mathrm{c.c}\Big]^4,
\label{eqn1}
\end{split}
\end{equation}
where $\xi_d = \xi_t +\xi_r = \frac{\tilde{\epsilon}}{2}\left(\frac{1}{\delta_t} + \frac{1}{\delta_r}\right)$ has contributions from directly driving the readout cavity, and from driving the transmon indirectly via the dispersive coupling. We expand the product and collect terms that are responsible for the charge-sideband transition of interest. The term corresponding to the $\ket{f0}-\ket{g1}$ transition with a target mode ($m$) is: 
\be 
\tilde{H}_{sb} = \alpha\beta_t^{2}\beta_m\beta_r\xi^{*}_d\left(\tilde{b}^{\dagger}_{m}\tilde{c}^2e^{i\left(\omega_d  + \omega_m - 2\omega_t\right) t} + \mathrm{c.c}\right).
\ee 
Accounting for the anharmonicity of the transmon and Stark shifts produced by the drive tone, we obtain $\omega_d = (2\omega_c - \omega_m) + 2\pi\left(\alpha + \delta\right)$ for the $\ket{f0}-\ket{g1}$ resonance condition, with $\delta = 2\alpha\beta_r^2\left(\beta_t^2-\beta_m^2\right)|\xi_d|^2$ being the Stark shift. The resulting SWAP rate is:
\bea 
g_{f0-g1} &=& \alpha\beta_c^{2}\beta_m\beta_r\xi_d = \frac{\tilde{\epsilon}\alpha g_m g_r}{2\Delta_m\Delta_r}\left[\frac{1}{\delta_t} + \frac{1}{\delta_r} \right] \nonumber\\
&\sim& \sqrt{\chi_m\chi_r}~(\xi_t + \xi_r)/2.
\eea 
The sideband transition rate scales as the geometric mean of the dispersive shift of the readout cavity and the target mode, and is linearly proportional to the drive displacement, $\xi_d = \xi_t + \xi_r$. We chose the frequencies of the transmon, the storage modes, and the readout resonator so that the drive term mostly acts on the transmon and is dominated by $\xi_t$.
\begin{figure}[t]
\includegraphics[width=0.5\textwidth]{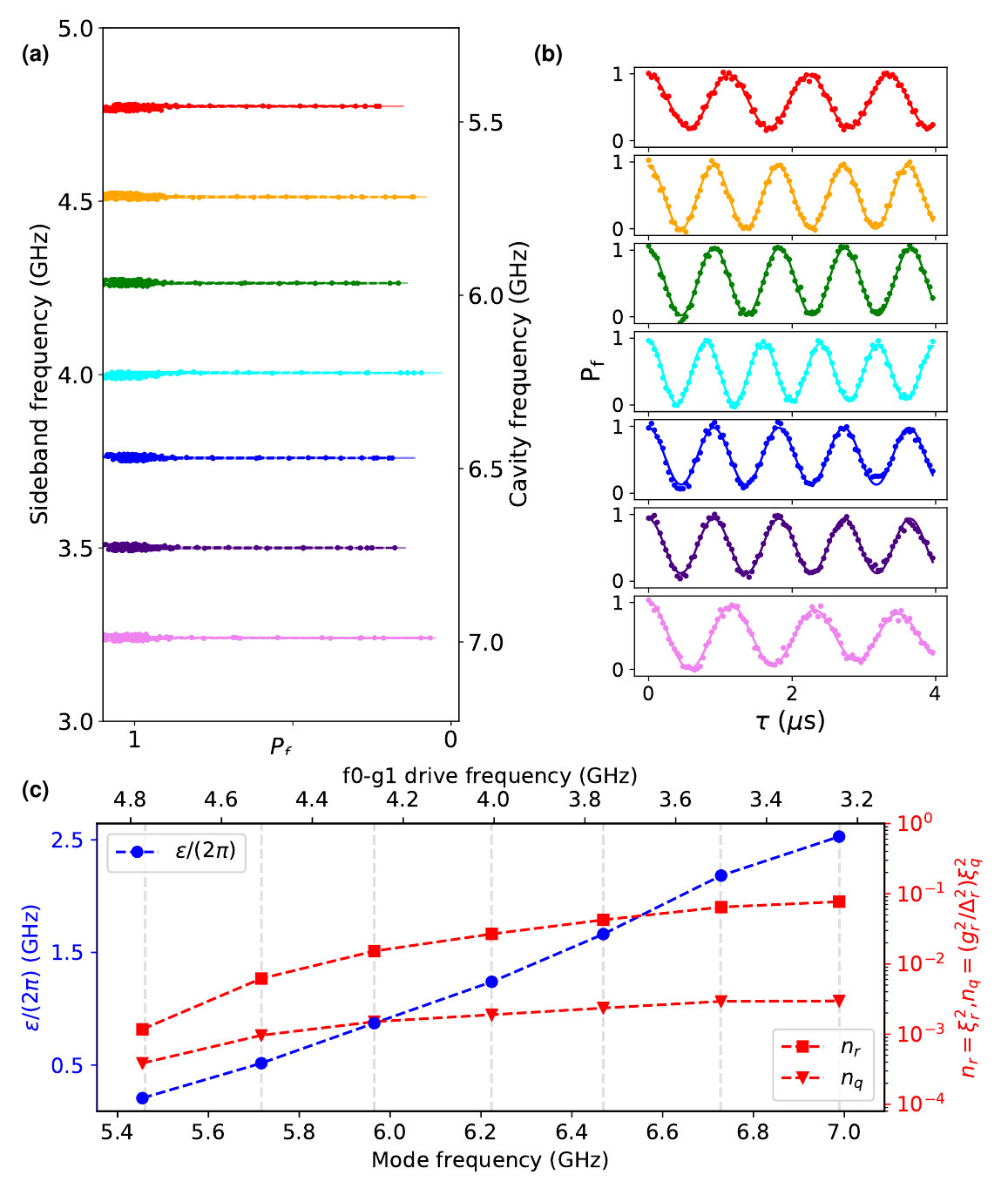}
\caption{(a) Spectroscopy of the $\ket{f0_i}-\ket{g1_i}$ transition for the first seven storage modes. The transmon is prepared in the $\ket{f}$ state, and a sideband drive tone is applied for a fixed duration at varying frequencies. The $\ket{f}$-state population is subsequently measured ($\ket{f}\rightarrow\ket{e}$).  The duration of the sideband pulse was chosen to correspond to that of a resonant $\pi$-pulse. (b) $\ket{f0_i}-\ket{g1_i}$ sideband Rabi oscillations for each of the modes, when driven on resonance for varying times. (c) (left) The cavity drive strength for each mode that results in a sideband Rabi rate of $g_{f0-g1} = 625$ kHz. The required cavity drive strength increases with mode frequency due to the larger detuning of the sideband frequency (top axis) from the transmon. (right) The expected off-resonant excitation of the readout ($n_r$) and the transmon ($n_q$). The increased dephasing of the sideband Rabi oscillations of the higher frequency storage modes is likely due to populating the readout cavity.}
	\label{sideband_spectra_and_oscillations}
\end{figure}
We find the $\ket{f0}-\ket{g1_m}$ transition with each through mode through sideband spectroscopy, the result of which is shown in SFig~\ref{sideband_spectra_and_oscillations}(a). The sideband drive amplitude and durations were adjusted to maximize the spectra contrast.  The corresponding Rabi oscillations when each $\ket{f0}-\ket{g1_m}$ transition is driven on resonance is shown in  SFig.~\ref{sideband_spectra_and_oscillations}(b). The sideband Rabi drive amplitude was adjusted across the different modes so as to obtain a $\ket{f0}-\ket{g1}$ $\pi$-pulse of $\sim 400-600$ ns. We plot the expected drive strength in the readout port required to realize a given sideband Rabi rate ($625$ kHz), and mean off-resonant occupation of the readout resonator ($n_r = |\xi_r|^{2}$) and the transmon ($n_t =(g_r^2/\Delta_r^2)|\xi_t|^{2}$), as a function of the mode frequency in SFig.~\ref{sideband_spectra_and_oscillations} (c).  The increased dephasing of modes at higher frequency is explained by the drive-induced excitation of the readout cavity. This effect can be greatly reduced by driving the storage modes or the transmon directly through ports that couple to them, instead of via the readout resonator.

\begin{figure*}[t]
  \begin{center}
    \includegraphics[width= 1.05\textwidth]{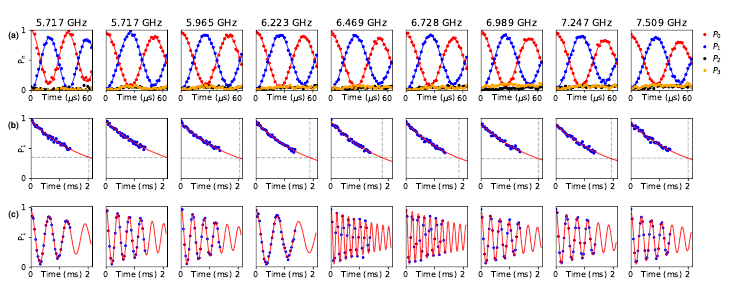}
    \caption{(a) Rabi oscillations in the $\ket{0}_i-\ket{1}_i$ subspace of each cavity mode induced by a weak resonant cavity drive ($\epsilon/2\pi$ = 10 kHz), while using a transmon drive to blockade $\ket{2}_i$. 
    (b) Lifetime measurements performed by loading each cavity mode in the $\ket{1}_{i}$ state using a $\pi$-pulse performed using the photon blockade cavity Rabi oscillation above.
    (c) Ramsey measurements of each mode performed using two photon blockade $\pi/2$-pulses with varying spacing, to take the cavity mode in and out of superposition. The population in a given Fock state ($P_n$) is measured using a narrow bandwidth transmon $\pi$-pulse at the corresponding photon number shifted transmon frequency. }
  	\label{cavity_coherence}
  \end{center}
\end{figure*}
\subsection{SNAP gates}
As described in the main text, we also perform universal cavity control operations using SNAP gates that involve cavity displacements and phase-gates implemented using photon-number selective transmon rotations. The gate is described by the following effective Hamiltonian rotating at the dressed transmon and storage mode frequencies:
\begin{equation}
\begin{split}
\hat{H} =~\chi\hat{a}^{\dagger}\hat{a}\ket{e}\bra{e} &+ \frac{\kappa}{2}\hat{a}^{\dagger}\hat{a}\left(\hat{a}^{\dagger}\hat{a}-1\right)  \\
&+ \left(\Omega(t)\ket{g}\bra{e} + \epsilon(t)\hat{a}+\mathrm{c.c.}\right),
\end{split}
\label{blockade_ham}
\end{equation}
where $\chi,\kappa$ are the dispersive and self-Kerr shifts of the target mode, and $\Omega, \epsilon$ represent the transmon and cavity drive strengths.  The control pulses (Fig.~3 (c) of the main text) were generated with the GrAPE algorithm~\cite{khaneja2005optimal}, using the package developed in \cite{leung2017speedup}. We perform the cavity displacements and the transmon rotations by driving the readout port at the transmon and storage mode frequencies respectively. The procedure followed to calibrate the drive amplitudes are as described in~\cite{chakramhe2020}.  
 
\subsection{Blockade Rabi oscillations}

 Photons can also be added to any of the storage modes by photon blockade as as shown in Fig.~3 (d) of the main text. Here, a resonant $\ket{g2_m}-\ket{e2_m}$ transmon drive results in the target mode ($m$) inheriting an anharmonicity equaling the rabi frequency of the transmon drive ($\Omega/2\pi \approx 107$ kHz). A weak cavity drive with strength $\epsilon \ll \Omega \ll  |\chi|$ subsequently induces a cavity Rabi oscillation in the  $\ket{0_m}-\ket{1_m}$ subspace of the target mode, allowing us to prepare an arbitrary single-qubit state in any mode. Cavity Rabi oscillations---induced by photon blockade of each of the storage modes---is shown in SFig.~\ref{cavity_coherence}(a).
 While the transmon is never directly occupied, the fidelity of the gates realized by photon blockade are still limited by transmon decay and dephasing. The infidelity arises through leakage to the dressed $\ket{2}$ state ($\epsilon/(\Omega^2 T^q)$), and Purcell decay of the $\ket{1}$ state ($\Omega^2/(\epsilon\chi^2 T^q)$) from transmon participation introduced by dressing from the blockade tone ($T_q = \mathrm{min}[T^1_q,T^2_q]$). Optimizing the cavity drive strength results in a minimum infidelity of $1/(\chi T^q)$ in the absence of cavity decay. These operations are additionally affected  by the intrinsic quality factors of the modes ($65-95\times10^6$), which results in an additional infidelity of $\sim 1-2~\%$. A more detailed description of the blockade gate, and its extensions for controlling qudits can be found in~\cite{chakramhe2020}. 

\subsection{Measuring cavity mode coherences}
\label{mode_coherence_with_qubit}
Each of the universal cavity control schemes previously described can be used to measure the coherence times of the cavity modes. The lifetime and Ramsey measurements providing the $T_1, T_2$'s of the storage modes of the multimode cavity [MM2(5N5)], Fig.~[4] of the main text), are shown in SFig.~\ref{cavity_coherence}(b) and (c). The cavity modes were prepared in the $\ket{1}$ and $\ket{0} + \ket{1}$ by using $\pi$ and $\pi/2$-pulses of the photon blockade induced cavity Rabi oscillations shown in SFig.~\ref{cavity_coherence}(a). We measure the same coherence times by using an $\ket{f0}-\ket{g1}$ sideband $\pi$-pulse to prepare the cavity state. The $T_2$ of the cavity modes range from $2-3$ ms. The additional pure dephasing responsible for the deviation of $T_2$ from $2T_1$ are from cavity frequency shifts induced by random transmon thermal excitations. Similar to dephasing of a qubit from thermal noise in a coupled oscillator~\cite{clerk2007using}, this is given by  $\Gamma_\phi =  \gamma\mathrm{Re}[\sqrt{(1+ i\chi/\gamma)^2 + 4i\chi n_{\mathrm{th}}^q/\gamma}-1]/2$, where $\gamma = 1/T_1^q$. Because the system is in the $\chi \gg \gamma$ regime, this reduces to $\Gamma_\phi = n_{\mathrm{th}}^q/T_1^q$ . The $T_2$ limits corresponding to the light blue bands in Fig.~4(b) of the main text, are obtained by using $n_{\mathrm{th}}^q = 1.2 \pm 0.5\%$ and $T_1^q = 86 \pm 6 \mu s$.  

\bibliography{thebibliography}